\documentstyle[prd,aps,psfig]{revtex} 

\begin{document}

\twocolumn

\def\beq{\begin{equation}}
\def\eeq{\end{equation}}
\def\bea{\begin{eqnarray}}
\def\eea{\end{eqnarray}}
\def\ben{\begin{enumerate}}
\def\een{\end{enumerate}}
\def\la{\langle}
\def\ra{\rangle}
\def\a{\alpha}
\def\b{\beta}
\def\g{\gamma}
\def\d{\delta}
\def\e{\epsilon}
\def\phi{\varphi}
\def\k{\kappa}
\def\l{\lambda}
\def\m{\mu}
\def\n{\nu}
\def\o{\omega}
\def\p{\pi}
\def\r{\rho}
\def\s{\sigma}
\def\t{\tau}
\def\L{{\cal L}}
\def\S{\Sigma }
\def\gsim{\; \raisebox{-.8ex}{$\stackrel{\textstyle >}{\sim}$}\;}
\def\lsim{\; \raisebox{-.8ex}{$\stackrel{\textstyle <}{\sim}$}\;}
\def\gtrsim{\gsim}
\def\lessim{\lsim}
\def\loc{{\rm local}}
\def\vm{v_{\rm max}}
\def\bh{\bar{h}}
\def\del{\partial}
\def\nab{\nabla}
\def\half{{\textstyle{\frac{1}{2}}}}
\def\fourth{{\textstyle{\frac{1}{4}}}}

\title{Generally covariant model of a scalar field with 
high frequency dispersion\\ 
and the cosmological horizon problem} 
 
\author{Ted Jacobson\thanks{jacobson@physics.umd.edu} 
and  David Mattingly\thanks{davemm@physics.umd.edu}}
\address{Department of Physics, University of Maryland,
College Park, MD 20742-4111, USA}
\maketitle

\begin{abstract}
Short distance structure of spacetime may show up in the form
of high frequency dispersion. Although such dispersion is not locally 
Lorentz invariant, we show in a scalar field model how it can nevertheless 
be incorporated into a generally covariant metric theory of gravity provided 
the locally preferred frame is dynamical. We evaluate the resulting 
energy-momentum tensor and compute its expectation value for a 
quantum field in a thermal state.  The equation of state differs
at high temperatures from the usual one, but not by enough
to impact the problems of a hot big bang cosmology. 
We show that a superluminal dispersion relation can solve the
horizon problem via superluminal equilibration, 
however it cannot do so 
while remaining outside the Planck regime unless the
dispersion relation is artificially chosen to have a
rather steep dependence on wavevector.
\end{abstract}


\section{Introduction}

Recent studies of the role of trans-Planckian physics in the
Hawking effect and in inflationary cosmology have exploited
scalar fields with high frequency dispersion as a model of 
how short distance physics might affect the behavior of quantum
fields in such settings. 
These models involve a preferred frame in which the 
distinction between high and low frequency is made. 
If such a matter 
field is to be coupled to the spacetime metric in a generally 
covariant theory of gravity, the preferred frame must be 
treated as a dynamical quantity rather than as a fixed background 
structure. One way of doing this was formulated by us in a 
previous paper\cite{aether}, in which the preferred frame
is determined by a dynamical unit timelike vector field $u^a$.
Using this formulation we obtained an expression 
for the stress tensor of a scalar field with dispersion. 

In the present paper we use this stress tensor  
to evaluate the equation of state in flat spacetime 
for a thermal state of the field.
This equation of state is then used in a fluid description 
of the matter field in a cosmological model, and the
implications for a hot big bang cosmology are examined.
We also investigate under what circumstances 
superluminal equilibration associated with high frequency
dispersion can solve the cosmological horizon problem.

In considering only the flat space thermal
state of the matter field we are adopting an
adiabatic approximation which precludes effects related to
out of equilibrium phenomena including particle creation 
in the dynamical background of a cosmological metric
such as were considered recently for fields with
high frequency dispersion in the context of 
inflation\cite{Martin,Niemeyer}.
This is a good approximation for frequencies larger than
the expansion rate $H$. It can therefore be used to 
study the effects of dispersion for frequencies of 
order $k_0$ provided $H\ll k_0$. 

\section{Model field theory}

Various possibilities exist for the kinetic 
terms in the action for $u^a$. For the purposes of
illustration in this paper we choose
the ``minimal theory" of
Ref. \cite{aether},
\bea
S_{min}[&&g_{ab},u^a,\l]=
\int d^4x\, \sqrt{-g}\, \nonumber\\
&&~~\Bigl(
-a_1 R 
- b_1\, F^{ab}F_{ab}
+\l(g_{ab}u^a u^b -1)
\Bigr),
\label{Smin}
\eea
where
\beq
F_{ab}:= 2\nabla_{[a} u_{b]}.
\eeq 
The field $\l$ is just a Lagrange multiplier whose variation
enforces the constraint that $u^a$ be a unit vector.

For the matter content we are interested in a scalar
field with high frequency dispersion
$\o^2=|\vec{k}|^2[1+g(|\vec{k}|/k_0)]$, where $g$ is a function
that vanishes at zero, and $k_0$ is a constant with the dimensions 
of inverse length which sets the scale for deviations from 
Lorentz invariance. 
It has been suggested that 
such a modified dispersion relation might arise 
in loop quantum gravity\cite{GP,MT}, 
or in string theory or other approaches to quantum
gravity,
or more generally from an unspecified modification
of the short distance structure of 
spacetime (see for example \cite{Amelino,river}).
Possible observational consequences have been
the subject of recent study 
(see for example \cite{Amelino,Bertolami,Bear,Carmona} and references
therein), and the role
of such dispersion in the Hawking process\cite{river}
and in the generation of inflationary 
primordial density fluctuations\cite{Martin,Niemeyer}
have been examined. 

Absent a reliable theory of such modifications, 
it makes sense simply to expand in $k/k_0$.
We consider here the lowest order modification 
for a scalar field that 
is invariant under rotations and 
analytic in $k$, which is given by 
\beq
\o(k)^2=|\vec{k}|^2 -|\vec{k}|^4/k_0^2,
\label{dr}
\eeq
where $k_0$ is a dimensionful parameter.
With $k_0^2>0$ this yields a subluminal group velocity,
and with $k_0^2<0$ it is superluminal.  The pathology
at $|\vec{k}|=k_0$ in the subluminal case is irrelevant 
since (\ref{dr}) is only regarded as the first terms 
in an expansion. When we evaluate the equation of state, 
we shall impose a cutoff that avoids this pathology.

The dispersion relation (\ref{dr}) 
can be produced by adding a term to the action
with four spatial derivatives. 
Previously this has been done in 1+1 dimensional models
(for a review see \cite{river}), and recently such models 
have been generalized to field theory in 
a background 3+1 dimensional Robertson-Walker spacetime\cite{Martin,Niemeyer}.
Here we extend these models to a general 3+1 dimensional spacetime.
This can be accomplished 
in a generally covariant manner, 
consistently with spatial rotation invariance,
with the action $S_\phi=\int d^4x\, \sqrt{-g}\,{\cal L}_\phi$,
where 
\beq
{\cal L}_\phi=\frac{1}{2}\Bigl(\nabla^a\phi \nabla_a\phi
+k_0^{-2}(D^2\phi)^2\Bigr).
\label{Lmod}
\eeq
Here $D^2$ is the covariant spatial Laplacian,
i.e.,
\beq
D^2\phi=-D^aD_a\phi
=-q^{ac}\nabla_a(q_c{}^b\nabla_b \phi), 
\eeq
where $D_a$ is the spatial covariant derivative 
operator\cite{wald} and $q_{ab}$ is the 
(positive definite) spatial metric orthogonal to 
the dynamical unit vector $u^a$,
\beq
q_{ab}:=-g_{ab}+u_au_b.
\label{q}
\eeq

The equation of motion for the metric takes the Einstein form
\beq
G_{ab}= 8\pi G\Bigl(T^{(u)}_{ab} + T^{(\phi)}_{ab}\Bigr),
\label{geq}
\eeq
where $G=1/16\pi a_1$, and
with
\beq
T^{(u)}_{ab}=
-4b_1
(F_{am}F_b{}^m-\fourth F^2 g_{ab}) 
+2\l u_au_b,
\label{tu}
\eeq
\bea
&&T^{(\phi)}_{ab}=
\nabla_a\phi\nabla_b\phi-
{\cal L}_\phi\, g_{ab}\nonumber\\
&&\!\!\!- k_0^{-2}\Bigl[
2 D^2\phi\, u^m u_{(a}\nab_{|m|}D_{b)}\phi
+2 \nab_m(D^2\phi\, q_{(a}{}^m)\nab_{b)}\phi\nonumber\\
&&~~~~~~~~-\nab^m(q_{ab}D^2\phi\, D_m\phi)
\Bigr].
\label{tdisp}
\eea
(The constraint equation $g_{ab}u^au^b=1$ has been used 
to drop the contribution to  
(\ref{tu}) that would have come from the variation
of $\sqrt{-g}$ in the constraint term of the action 
(\ref{Smin}).)

The equation of motion for the field $u^a$ 
takes the form
\beq
\nab^b F_{ba}=-\frac{1}{2b_1}
\Bigl(\l u_a + \frac{1}{2}\frac{\d S_{\phi}}{\d u^a}\Bigr).
\label{ueqmatter}
\eeq
with
\bea
&&\frac{\d S_\phi}{\d u^a}=
2k_0^{-2}u^b\nonumber\\
&&\Bigl[
D^2\phi\, \nab_{(a}\bigl(q_{b)}{}^m\nab_m \phi\bigr)
-\nab_m\bigl(D^2\phi\, q_{(a}{}^m\bigr)\nab_{b)}\phi\Bigr].
\label{dsdu}
\eea
Contracting (\ref{ueqmatter}) with $u^a$ we obtain an
expression for $\l$ in terms of the other fields:
\beq
\l=2b_1 u^a \nab^b F_{ab}-\frac{1}{2}u^a\frac{\d S_{\phi}}{\d u^a},
\label{l}
\eeq
where, from (\ref{dsdu}),
\beq
u^a\frac{\d S_\phi}{\d u^a}= 
4k_0^{-2}u^au^b
D^2\phi\, \nab_{[m}\phi\, \nab_{a]}q_b{}^m.
\label{udsdu}
\eeq

\section{Thermal state in a Robertson-Walker cosmology}

Let us now specialize to a
Robertson-Walker (RW) spacetime, 
in the semiclassical framework where the metric
and $u^a$ are treated as classical fields and the scalar
field is a quantum field (which therefore has a well-defined
thermal equilibrium state). In the field equations for the
metric and for $u^a$ we take the expectation value of the 
$\phi$-terms. Assuming $u^a$ shares the RW
symmetry it must be the cosmological rest frame.
The tensor $F_{ab}$ then vanishes, so $\l$ is just
determined by the matter term in (\ref{l}). (If there are further
terms involving $u^a$ in the action then there are additional
contributions to $\l$.) 

With RW symmetry we have $\nab_a u_b = H q_{ab}$, where 
$H=\dot{a}/a$ is the usual Hubble ``constant". 
Iterating this identity we find  
$\nabla_a q_{mn} = H(q_{am}\, u_n + q_{an}\, u_m)$,
from which it follows that the contraction
(\ref{udsdu}) is given by 
\beq
u^a\frac{\d S}{\d u^a}= 
6k_0^{-2}\, H\, \dot{\phi}\, D^2\phi,
\label{udsduRW}
\eeq
where $\dot{\phi}=u^m\nab_m\phi$. 

Suppose now that the scalar field is well 
approximated by an adiabatically evolving 
thermal state.
This would be the case if (i) there are interactions
that produce an equilibration rate which is large 
compared to the expansion rate, and (ii) the thermal 
frequency is also large compared to the expansion
rate. In this case the expectation value 
of (the Hermitian part of) the operator $\dot{\phi}\, D^2\phi$
vanishes, since the operator is odd under time reversal 
while a thermal state is invariant. Using (\ref{l}) and
(\ref{udsduRW}) this implies
that, in the minimal model, $\la \l\ra=0$. 
(If other terms are included in the 
action for $u^a$ then $\la\l\ra$ would not vanish, although it
would still not receive contributions from this matter field.)
Therefore in this model $\la T^{(u)}_{ab}\ra=0$, so the
only contribution to the cosmological stress tensor comes from 
the scalar field.

\subsection{Thermal equation of state}
Consider now the expectation value $\la T^{(\phi)}_{ab}\ra$ of the 
stress tensor (\ref{tdisp})---or more precisely of
its Hermitian part---in a thermal state.
Note first that time reversal
invariance of the thermal state requires an even number of time 
derivatives of $\phi$, and spatial isotropy requires an even number of 
spatial derivatives of $\phi$, 
in order for the thermal expectation value not to vanish. Thus terms with
an odd number of derivatives of $\phi$ do not contribute. 
Let us call this the ``odd derivative rule".
We can use this rule to see that 
the expectation value $\la{\cal L}_\phi\ra$ vanishes. Integrating
by parts, ${\cal L}_\phi$ can be expressed as 
a term that vanishes since $\phi$ satisfies its equation
of motion, plus the total derivative of 
an expression involving only terms with an odd number of derivatives
of $\phi$. The total derivative can be taken out of the expectation
value, hence that term vanishes.

The part of (\ref{tdisp}) multipled by $k_0^{-2}$ has three terms.
The expectation value of the first term has a single time derivative,
hence vanishes by time reversal symmetry. The expectation value of the
third term is the gradient of an expression with three spatial derivatives
which vanishes by the odd derivative rule. 
The second term can be integrated by 
parts, and the resulting total derivative piece has vanishing expectation
value by the odd derivative rule, which leaves only
$2q_{(a}{}^m\la D^2\phi\,\nab_m\nab_{b)}\phi\ra$. In a thermal state
this must have the form $Au_au_b + Bq_{ab}$, where $u^a$ is the rest frame
defined by the thermal bath. To find $A$ we contract with
$u^au^b$, which yields $A=0$ due to the factor $q_{a}{}^m$. 
To find $B$ we contract with $q^{ab}$ and divide by 3, hence this
term contributes $-\frac{2}{3}\la D^2\phi\, D^2\phi\ra$. The 
expectation value of the stress tensor thus takes the form 
\beq
\la T_{ab}\ra = \rho\, u_a u_b + P\, q_{ab},
\eeq
with energy density $\rho$ and pressure $P$ given by 
\bea
\rho&=&
\la\dot{\phi}^2 \ra\label{rho}\\
P &=&\la{\textstyle\frac{1}{3}} (D\phi)^2 
-{\textstyle\frac{2}{3}}k_0^{-2} (D^2\phi)^2\ra.\label{P}
\eea

To evaluate the density and pressure we expand the field
in Fourier components using the dispersion relation
(\ref{dr}) and sum the contributions from 
the modes, weighting each by the thermal expectation
value of the number operator $e^{\o(\vec{k})/T} -1$, 
which yields
\beq
\r=\int \frac{d^3k}{2\pi^3\o(\vec{k})} \frac{\o(\vec{k})^2}
{e^{\o(\vec{k})/T} -1},
\eeq
\label{rint}
\beq
P=\frac{1}{3}\int \frac{d^3k}{2\pi^3\o(\vec{k})} 
\frac{\o(\vec{k})^2-|\vec{k}|^4/k_0^2}{e^{\o(\vec{k})/T} -1}.
\label{pint}
\eeq
For low temperatures $T\ll k_0$, only modes with $k\ll k_0$
contribute significantly, hence we recover the standard result
for massless radiation, $P=\frac{1}{3}\rho$, and the 
energy density scales as $\rho\propto T^4$.

Before looking at the exact temperature dependence,
let us consider the high temperature limit.
The nature of this limit depends on whether the dispersion is
sub- or super-luminal. In the superluminal
case $k_0^2<0$, we can sensibly use the dispersion relation (\ref{dr})
out to arbitrarily large $k$, 
so in the high temperature limit
only wave vectors $|\vec{k}|\gg |k_0|$ are relevant. 
For such wave vectors we have 
$\o(\vec{k})^2\simeq |\vec{k}|^4/|k_0^2|$, hence
the energy density scales as $\rho\propto T^{5/2}k_0^{3/2}$,
and $P\simeq \frac{2}{3}\rho$. 
This should be taken only as a qualitative indication
of what might occur,
since one would expect that for $|\vec{k}|\gsim k_0$ 
further terms in the $k$-expansion
of the dispersion relation become important.

In the subluminal case $k_0^2>0$ we must impose a cutoff so as not
to enter the unphysical region where the dispersion relation would 
yield $\o(\vec{k})^2<0$. We choose to impose the cutoff 
at $|\vec{k}|=k_0/\sqrt{2}$, where $\o(\vec{k})$ attains its
maximum value. This is similar to a lattice cutoff
for which, in one dimension, the dispersion relation is 
$\o(k)=(2/\d)\sin k\d/2$, and for which $|k|\le \pi/\d$
exhausts all the independent modes. With this cutoff in place,
$\o(\vec{k})/T\ll1$ for all modes in the high temperature limit, hence the
expectation value of the number operator tends toward $T/\o(\vec{k})$.
Thus the energy density scales as  $\rho\propto T k_0^3$ in this limit.
As for the equation of state, the ratio of the integrals 
yields $P/\rho\simeq 0.174$.

\begin{figure}[]
\centerline{
\psfig{figure=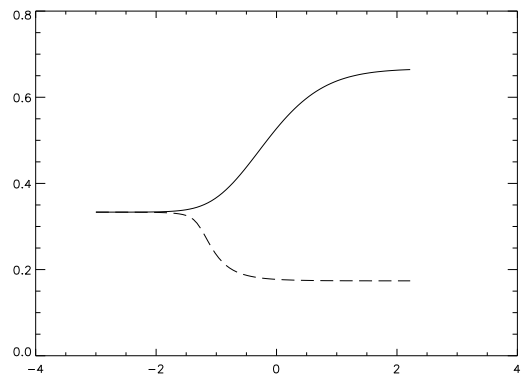,angle=0,height=5.5cm}
}
\vskip 2mm
\caption{\small Thermal expectation values for $P/\r$ 
vs. $\log(T/k_0)$ in the presence of high
frequency dispersion. The dashed and solid curves are for the
 sub- and superluminal cases respectively.} 
\label{eos}
\end{figure}

The interpolation between the low and high temperature limits can 
be determined by a numerical calculation of the integrals.
In Fig. \ref{eos} we plot $P/\rho$ vs. $\log_{10}(T/k_0)$.
It is seen that the equation of state smoothly connects the
low and high temperature limits. For the subluminal case most of 
the interpolation takes place over the range of temperatures
$10^{-1.5}<T/k_0<10^{-0.5}$, while in the superluminal 
case most of the interpolation takes place over the somewhat
higher range $10^{-1}<T/k_0<10$. 

\subsection{Cosmological implications}
The modified equation of state derived here can 
be used in a cosmological model where thermal matter with 
non-Lorentz-invariant dispersion acts as a source of gravity.
The Bianchi
identity implies the energy-momentum tensor is divergence-free,
which for a perfect fluid implies $\dot{\r}=-3H(\r+p)$.
Together with the equation of state this determines the evolution of 
the temperature. In the
superluminal case local energy-momentum conservation is also
independently implied by the matter field equations as usual, 
however in the subluminal 
case the dynamics is not self-contained because of the high
wave vector cutoff. If the cutoff is imposed at a fixed proper
wave vector, new modes are added as the universe expands, and 
the field dynamics does not specify into which state these
modes are born. The assumptions of a thermal stress tensor and 
of energy-momentum conservation require that, whatever their birth state, 
they rapidly equilibrate with the rest of the modes. This is reasonable
under the assumption that the system is in the adiabatic regime.

It is interesting to ask whether high frequency dispersion 
could have any impact on the cosmological problems that led to 
the invention of the inflationary scenario. In particular, could 
it provide a solution of these problems not requiring inflation?
The modification of the equation of state caused by the 
dispersion would affect the quantitative details
of the evolution of the scale factor at times when the
typical wavevectors are of order $k_0$ or greater, which is 
presumably only in the very early universe. However, as we have seen,
the equation of state only changes from $P=\r/3$ to $P=(2/3)\r$ in the
superluminal case and to $P=0.17\r$ in the subluminal case.
Neither are significant enough to qualitatively change the dynamics 
or horizon size.  

\subsection{Superluminal equilibration and the horizon problem}
In the superluminal case, the fact that influences travel faster than
light at wavevectors $k\gsim k_0$ opens up the possibility of 
solving the horizon problem via superluminal equilibration.
The coordinate distance covered by a wavepacket with proper group
velocity $v_g$ is $\Delta x=\int v_g dt/a$.
For a dispersion relation that goes as $\o\sim k^n$ at large
wave vectors the group velocity goes as $v_g\sim k^{n-1}$.
It is easy to show that the typical wavevector at the peak of
a thermal distribution scales as $a^{-1}$ (as long as the 
dispersion relation is homogeneous), hence the typical group
velocity scales as $a^{1-n}$, so we have $\Delta x\sim\int dt/a^n$.
If this diverges then the horizon problem is solved assuming the
framework of this model.

With the above dispersion relation, the equation of state
is $P=(n/3)\r$, which according to the Einstein equation yields
the evolution of the scale factor $a(t) \propto t^{2/(n+3)}$.
Thus $\Delta x\sim\int dt\, t^{-2n/(n+3)}$, which diverges at
the lower limit for any $n\ge3$. 

Unfortunately it does not make much sense to view this
as a solution to the horizon problem, because most of the
$\Delta x$ is traversed during a regime in which the 
typical wavevector and energy density are so much larger 
than the Planck scale that we have no reason to trust the
semiclassical model at all. 
In particular, we have checked that in order 
for $\Delta x$ to surpass the horizon size, the evolution
must be extrapolated all the way back to a time at which 
the typical wavevector is roughly $k_0$ times
$[10^{57}(k_0/k_{Planck})^2]^{1/(n-3)}$. Presuming that $k_0$
is within a few orders of magnitude of the Planck scale, this 
typical wave vector exceeds the Planck scale unless $n\gsim50$.
Since we have no theory that would determine the maximum 
exponent $n$ appearing in a superluminal dispersion relation,
it would seem artificial at this stage to adjust $n$ in order to 
achieve a sub-Planckian resolution of the horizon problem.

\section*{Acknowledgements}

This work was supported in part by the National Science Foundation
under grant No. PHY98-00967.

\end{document}